\documentclass[twocolumn,showpacs,prl,floatfix]{revtex4}

\usepackage{graphicx}
\usepackage{dcolumn}
\usepackage{bm}

\usepackage{epsfig}

\def\be{\begin{equation}}
\def\ee{\end{equation}}
\def\bes{\begin{equation*}}
\def\ees{\end{equation*}}
\def\bea{\begin{eqnarray}}
\def\eea{\end{eqnarray}}
\def\beas{\begin{eqnarray*}}
\def\eeas{\end{eqnarray*}}

\begin{document}


\title{Vortex Lattices in Rotating Atomic Bose Gases with
Dipolar Interactions}

\author{N. R. Cooper$^{1}$, E. H. Rezayi$^{2}$ and S. H. Simon$^{3}$}

\affiliation{$^1$T.C.M. Group, Cavendish Laboratory, Madingley Road, Cambridge, CB3 0HE,
United Kingdom.\\
$^2$Department of Physics, California State University, Los Angeles, California 90032.\\
$^3$Lucent Technologies, Bell Laboratories, 600 Mountain View Avenue, Murray Hill, New Jersey 07974.}

\date{May 31, 2005}

\begin{abstract}

We show that dipolar interactions have dramatic effects on the
groundstates of rotating atomic Bose gases in the weak interaction
limit.  With increasing dipolar interaction (relative to the net
contact interaction), the mean-field, or high filling fraction,
groundstate undergoes a series of transitions between vortex lattices
of different symmetries: triangular, square, ``stripe'', and
``bubble'' phases.  We also study the effects of dipolar interactions
on the quantum fluids at low filling fractions. We show that the
incompressible Laughlin state at filling fraction $\nu=1/2$ is
replaced by compressible stripe and bubble phases.

\end{abstract}


\maketitle

Ultra-cold atomic Bose gases have emerged as remarkable systems with
which to study the unusual response of Bose-condensed systems to
rotation.  Experiments have allowed the imaging of lattices of
quantised vortices\cite{MadisonCWD00add}, and the study of their
collective dynamics\cite{coddington:100402}. New vortex lattice
structures in two-component condensates have been
observed\cite{schweikhard:210403}, and a novel regime of vortex
density in which the vortex cores overlap\cite{wgs} has been
accessed\cite{schweikhard:040404}.  Theory shows that, at very high
vortex density\cite{cwg}, atomic Bose gases should undergo a
transition into novel uncondensed phases closely related to the
incompressible liquids of the fractional quantum Hall
effect\cite{wgs,WilkinG00,CooperWadd,cwg}.

Although Bose condensation can occur for a non-interacting Bose gas,
the formation of arrays of quantised vortices under rotation relies on
non-vanishing (repulsive) interparticle interactions. In typical
atomic Bose condensates, the interactions are so short-ranged that
they can be viewed as local (contact) interactions.  However,
significant additional {\it non-local} interactions can arise if the
atoms have intrinsic or induced electric or magnetic dipole
moments\cite{MarinescuY98add,BaranovPS}.  The recent
achievement\cite{GriesmaierWHSP05} of the Bose condensation of
chromium (which has a large permanent magnetic dipole moment) has
opened the door to the experimental study of dipolar-interacting Bose
gases.  It is important to ask what are the effects of non-local
interactions on the properties of the vortex lattices.

In this paper, we study the effects of dipolar interactions on the
groundstate of a weakly interacting atomic Bose gas under rotation.
We show that the additional non-local interaction leads to dramatic
changes in the nature of the groundstate. Within mean-field theory,
the triangular vortex lattice predicted for contact interactions is
replaced by vortex lattices of different symmetries: square,
``stripe'' and ``bubble'' phases.  Furthermore, we study the
properties at very high vortex density where the groundstates for
contact interactions are incompressible quantum fluids.  We show that
with increasing dipolar interactions the Laughlin state, which is the
groundstate for pure contact interactions at filling fraction
$\nu=1/2$\cite{wgs}, is replaced by {\it compressible} states that are
well described by stripe and bubble phases.

\begin{figure*}
\centering
\includegraphics[width=17cm]{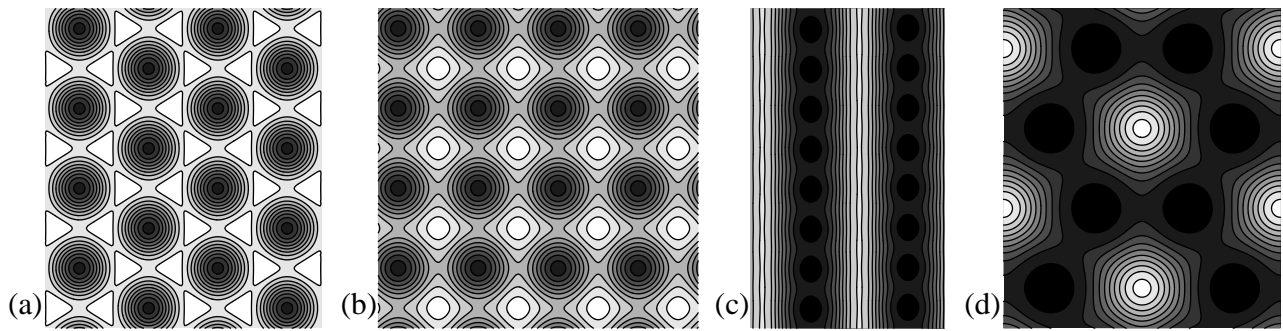}

\caption{\label{fig:images} Contour plots of the particle densities
for condensed states on a torus with $N_V=16$ vortices. The light
(dark) shading indicates high (low) particle density (on arbitrary
scales).  (a) Triangular vortex lattice ($a/b=\sqrt{3}/2$); (b) Square
vortex lattice ($a/b=1.0$); (c) ``Stripe'' phase (shown for $\alpha
=0.528$, $a/b=0.608$); (d) $q=4$ ``bubble'' phase ($a/b=\sqrt{3}/2$).}

\end{figure*}

We consider a system of bosonic atoms of mass $M$ confined to a
harmonic trap with cylindrical symmetry about the $z$-axis.  We denote
the natural frequencies of the trap by $\omega_\parallel$ and
$\omega_\perp$ in the axial and transverse directions, and the
associated trap lengths by $a_{\parallel,\perp}\equiv
\sqrt{{\hbar}/({M\omega_{\parallel,\perp}})}$.
The particles are taken to interact through both contact interactions
and additional dipolar interactions.  We consider the atomic dipole
moments to be aligned with the $z$-axis, as in the experiments
reported in Ref.\cite{GriesmaierWHSP05}, such that the net two-body
interaction is \be
\label{eq:dipolar}
V({\bm r}) = \frac{4\pi\hbar^2 a_s}{M} \,\delta^3(\bm{r}) +
C_d\frac{x^2+y^2-2z^2}{(x^2+y^2+z^2)^{5/2}} \ee where $a_s$ is the
$s$-wave scattering length arising from contact interactions\cite{as},
and $C_d$ is a measure of the strength of the dipolar interactions.
 
We study the regime of weak interactions, when the mean interaction
energy per particle is small compared to the trap energies,
$\hbar\omega_{\perp,\parallel}$. The single particle states are then
restricted to the groundstate of the trap in the $z$-direction (2D)
and to the lowest Landau level (LLL) of the $x-y$
motion\cite{wgs,ButtsR99}.  While non-rotating gases are typically far
from the weak interaction limit, this limit can be approached at high
angular momentum owing to the reduction in particle density through
the centrifugal spreading of the cloud\cite{schweikhard:040404}.
In this limit the interactions are fully specified by the Haldane
pseudopotentials\cite{haldanehierarchy}, $V_m$: the interaction energy
of a pair of particles with relative angular momentum $m$.  For
bosons, symmetry of the wavefunction means that only even $m$ contribute. The precise form of the dipolar
pseudopotentials depends in a non-trivial way on the trap asymmetry
$a_\parallel/a_\perp$.  For simplicity of presentation, we study the
limit $a_\parallel/a_\perp \to 0$, in which the dipole forces fall off
most quickly with increasing distance.
To leading order in $a_\parallel/a_\perp$, we find \bea V_0 & = &
\sqrt{\frac{2}{\pi}}\frac{\hbar^2 a_s}{Ma_\perp^2a_\parallel} +
\sqrt{\frac{2}{\pi}}\frac{C_d}{a_\perp^2a_\parallel} -
\sqrt{\frac{\pi}{2}}\frac{C_d}{a_\perp^3}\\ V_{m> 0} & = &
\sqrt{\frac{\pi}{2}} \frac{(2m-3)!!}{m!\,2^m}\frac{C_d}{a_\perp^3}
\eea $V_0$ represents the net {\it local} interaction, which has
contributions from both the contact and dipolar interactions;
$V_{m>0}$ represent {\it non-local} interactions which arise only from
the dipolar interaction.  We consider the relative sizes of the
non-local and local interactions to be variable, either by tuning the
dipolar interaction, $C_d$, or by tuning $a_s$ close to a Feshbach
resonance\cite{werner:183201} (which can allow $a_s$ to become
negative).
 We quantify their relative size by the ratio \be\alpha \equiv
\frac{V_2}{V_0} \ee 
We are
interested in  bulk properties, so perform calculations on a
periodic rectangular geometry (a torus) which accurately describes the
centre of an atomic gas where the particle density is approximately
uniform.  In the weak interaction limit the number density of vortices
is $n_V= 1/(\pi a_\perp^2)$, so a torus with sides $a$ and $b$
contains $N_V = ab/(\pi a_\perp^2)$ vortices.

First, we describe the mean-field groundstates. Within
Gross-Pitaevskii (GP) mean-field theory the groundstate is assumed to
be fully condensed, with condensate wavefunction, $\psi({\bm r})$.  In
the weak-interaction limit, $\psi(\bm{r})$ is found by minimising the
mean interaction energy for fixed average particle density, with the
wavefunction constrained to states in the 2D LLL\cite{ButtsR99}.
The problem is mathematically equivalent to the Ginzburg-Landau model
for a type-II superconductor close to $H_{c2}$\cite{Abrikosov57}
generalised to a non-local interaction. A phenomenological model of
this kind has been discussed in the context of
superconductivity\cite{YeoM97}, but we are not aware of general
solutions for the vortex lattice groundstates. 
 We have found the
mean-field groundstates for the interactions (\ref{eq:dipolar}) by
numerical minimisation on a torus with up to $N_V=24$ vortices ({\it
i.e.}  exploring periodic states with up to 24 vortices in the unit
cell).  As a function of $\alpha$, the groundstate undergoes a series
of transitions between states of different translational symmetries.
Representative images of the particle
distributions in these states are shown in
Fig.\ref{fig:images}\cite{symmetry}.

The groundstates we find are: a triangular lattice of single vortices
($0 \leq \alpha \leq 0.20$); a square lattice of single vortices
($0.20\leq \alpha \leq 0.24$); a stripe phase ($0.24\leq \alpha \leq
0.60$)\cite{narrow}.  The stripe phase consists of broad
lines of high particle density separated by rows of closely-spaced
vortices. The vortices are ordered along the rows, so the state is a
``stripe crystal'' with crystalline order in both directions.
For $\alpha\geq 0.60$ the states consist of clusters of high particle
density arranged in a triangular lattice.  Owing to the similarity to
crystalline states of electrons in high Landau
levels\cite{FoglerKS96add}, we refer to these as ``bubble'' states. We
find a sequence of bubble states, which we label by the number $q$ of
vortices associated to each bubble.
The bubble states we find as groundstates are $q=4$ ($0.60\leq
\alpha\leq 0.91$), $q=5$ ($0.91\leq \alpha \leq 1.4$), $q=6$ ($1.4\leq
\alpha\leq 2.0 $), $q=7$ ($2.0\leq \alpha \leq 2.7$), $q=8$ ($2.7\leq
\alpha$).  The particle distributions of the bubble states with $q\geq
4$ resemble Fig.\ref{fig:images}(d), with additional vortices confined
to the honeycomb network separating the bubbles of particles.
We have not looked for states with $q> 8$, but expect that
states of arbitrarily large $q$ will appear.

\begin{figure*}

\includegraphics[width=17.5cm]{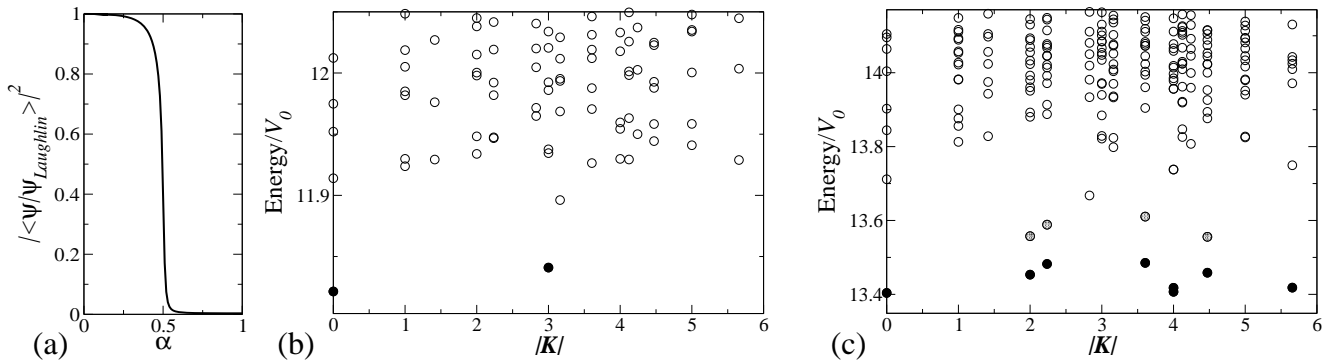}

\caption{\label{fig:half} Results of exact diagonalisation studies at
$\nu=1/2$. (a) Overlap of the exact groundstate with the Laughlin
state ($N=8$, $N_V=16$, torus aspect ratio $a/b=1$).  (b) Energy
spectrum for $\alpha = 0.528$ ($N=9$, $N_V=18$, $a/b=0.82$).  The
filled circles are for momenta ${\bm K}=(0,0)$ and $(0,3)$, showing
evidence for a stripe state.  (c) Energy spectrum at $\alpha = 0.758$
($N=8$, $N_V=16$, $a/b=\sqrt{3}/2$).  The filled circles are at ${\bm
K}=(0,0), (4,0), (0,4), (4,4), (2,0), (1,2), (3,2), (2,4)$, which are
the reciprocal lattice vectors of the $q=4$ bubble state.  The shaded
symbols are at ${\bm K}=(0,2), (2,1), (2,3), (4,2)$, corresponding to
the reciprocal lattice vectors of a rotated crystal.}

\end{figure*}

We believe that it should be possible experimentally to access these
new vortex lattice groundstates. To make contact with experimental
parameters, we now examine the case of a spherically symmetric trap
($a_\parallel = a_\perp$). In this case  the triangular
lattice is replaced by new groundstates if $a_s$ is
tuned\cite{werner:183201} to values $a_s \alt -0.13 C_d M/\hbar^2$.
Note that, despite this negative value of $a_s$ the contribution of
the dipolar interaction makes $V_0$ positive, which, in the weak
interaction limit, is sufficient to ensure stability  to
collapse.
Under these circumstances, for a mean particle density $n_{3d}$, the
interaction energy per particle is of order $C_d n_{3d}$, so the weak
interaction limit requires $n_{3d} \lesssim \hbar\omega_\perp/C_d$.
Choosing the value of $C_d$ appropriate for
chromium\cite{GriesmaierWHSP05}, which has a magnetic dipole moment of
6 Bohr magnetons, and taking $\omega_\perp = 2\pi\times 100 \mbox{rad
s}^{-1}$, the weak interaction limit should be a good approximation
for densities less than about $10^{14}\mbox{cm}^{-3}$.

We now turn to discuss the groundstates beyond the mean-field
approximation.  In Ref.\cite{cwg} it was shown that the parameter
controlling the validity of mean-field theory for a rotating atomic
Bose gas is the filling fraction, $\nu \equiv {n}/{n_V}$, where $n$
and $n_V$ are the number densities (per unit area) of particles and
vortices.  For contact interactions, the triangular vortex lattice
predicted by mean-field theory was shown to be destroyed by quantum
fluctuations for $\nu < \nu^{\rm tri}_c$, with $\nu^{\rm tri}_c\sim
6$\cite{cwg}, and replaced by new groundstates which include
incompressible liquids closely related to fractional quantum Hall
states\cite{wgs,WilkinG00,CooperWadd,cwg}.

We have studied the effects of dipolar interactions on these
strongly-correlated groundstates using exact diagonalisation studies.
We now return to the case of $a_\parallel/a_\perp \rightarrow 0$ and
focus our attention on the filling fraction $\nu=1/2$, for which the
exact groundstate for contact interactions ($\alpha=0$) is the
$\nu=1/2$ bosonic Laughlin state\cite{wgs}.  This is the dominant
incompressible state of rotating bosons with contact interactions.
We find that the exact groundstate in the presence of dipolar
interactions is well described by the Laughlin state up to
$\alpha \simeq 0.5$. At this point there is an abrupt transition in
the groundstate and its overlap with the Laughlin state falls to a
very small value [Fig.\ref{fig:half}(a)].  Our studies show that the
states that replace the Laughlin state at $\alpha \gtrsim 0.5$ are
{\it compressible} phases which are well described by the mean-field
groundstates discussed above.  Here we present evidence of a stripe
phase at $\alpha = 0.528$ and of the $q=4$ bubble phase at $\alpha =
0.758$. (For larger $\alpha$ we find evidence of bubble phases with
larger $q$.)  To identify these broken-symmetry states we make
extensive use of the classification of energy eigenstates by a
conserved momentum\cite{haldanemtm}.  We express the momentum as a
dimensionless vector ${\bm K} = (K_x,K_y)$ using units of $2\pi\hbar/a$ and
$2\pi\hbar/b$ for the $x$ and $y$ components, and report only positive
$K_x,K_y$ [states at $(\pm K_x,\pm K_y)$ are degenerate by
symmetry]. We identify the broken translational symmetry of the
groundstates by making use of the fact that this leads to the
appearance of quasi-degeneracies in the spectrum at wavevectors equal
to the reciprocal lattice vectors of the crystalline
order\cite{RezayiHY99add}.

Evidence of stripe ordering at $\alpha = 0.528$ is presented in
Fig.\ref{fig:half}(b). This shows the excitation spectrum on a torus
with an aspect ratio chosen to be consistent with the mean-field
groundstate.  The quasi-degeneracy of the groundstate at ${\bm
K}=(0,0)$ with a state at $(0,3)$ indicates a strong tendency to
translational symmetry breaking in the stripe pattern found in
mean-field theory: three stripes lying parallel to the short axis of
the torus.  However, unlike the mean-field state, we find no evidence
of crystalline order parallel to the stripes: the groundstate appears
to be a ``smectic''\cite{ChaikinLadd} in which translational symmetry
is broken in only one direction.

Evidence for the formation of the $q=4$ bubble phase is shown in
Fig.~\ref{fig:half}(c).  At filling fraction $\nu=1/2$, each bubble
contains $\nu q = 2$ particles, so this state can also be described as
a triangular lattice of pairs of particles.  The low energy states
shown in Fig.~\ref{fig:half}(c) as filled symbols correspond to
wavevectors at the reciprocal lattice vectors of the $q=4$ bubble
state [shown in Fig.\ref{fig:images}(d)].  Athough these states do not
seem to be cleanly separated from the rest of the spectrum, the other
low energy states (shaded symbols) can be understood as arising from a
nearby (in energy) crystalline configuration in which the crystal is
rotated through $90^\circ$.

The appearance of stripe and bubble states at $\nu=1/2$ indicates that
these states are much more stable to quantum fluctuations than is the
triangular vortex lattice (which is unstable for $\nu < \nu_c^{\rm
tri} \sim 6$\cite{cwg}).  This enhanced stability can be understood
within a simple Lindemann analysis, in which one asserts that quantum
melting occurs when the quantum fluctuations of a lattice site exceed
a multiple $c_L$ of the lattice spacing. 
Treating the fluctuations of each vortex independently, one
expects the triangular vortex
lattice to melt for $\nu < \nu_c^{\rm tri} =
{\sqrt{3}}/{{(2\pi c_L^2)}}$\cite{cwg,RozhkovS96}.  For the square
lattice, we find $ \nu^{\rm sq}_c = {1}/{{(\pi c_L^2)}} =
({2}/{\sqrt 3})\nu_c^{\rm tri}$, close to that for the triangular
lattice.
The enhanced stabilities of the stripe and bubble phases arise from
the existence of larger lengthscales and larger numbers of particles
per unit cell (as compared to the triangular or square
vortex lattices).  For the $q$ bubble phase, the bubbles form a
triangular lattice with lattice constant $\sqrt{ q
\frac{2\pi}{\sqrt{3}}} a_\perp$.  Applying the Lindemann analysis to
the quantum fluctuations of the centre-of-mass of a bubble, we find $
 \nu^{q}_c = \frac{\sqrt{3}}{2\pi c_L^2}\frac{1}{q^2} = \nu_c^{\rm
tri}/q^2$. Even for the smallest ($q=4$) bubble state, this critical
filling fraction is very much smaller than that for the triangular
lattice.  For the stripe phases, there are two lengthscales: the
inter-vortex separation in the directions parallel, $R_\parallel$, and
perpendicular, $R_\perp$, to the stripes. One therefore expects two
transitions: when fluctuations of the vortices along the stripes
exceed $c_L R_\parallel$ there is a loss of order in that direction,
leading to a smectic phase; when fluctuations perpendicular to the
stripes exceed $c_L R_\perp$ the stripe ordering will finally be lost.
Assuming that, for this final transition, of order
$R_\perp/R_\parallel$ vortices must fluctuate together, we find that
loss of stripe order should occur at $\nu_{c,\perp}^{\rm stripe} =
\frac{1}{\pi c_L^2}\left(\frac{R_\parallel}{R_\perp}\right)^2 =
({2}/{\sqrt{3}})\left({R_\parallel}/{R_\perp}\right)^2 \nu_c^{\rm
tri}$. Over the range $\alpha = 0.24-0.60$ for which the stripe is the
mean-field groundstate the ratio $R_\perp/R_\parallel = 1 - 2.59$, so
the critical filling fraction for the stripe can be as small as
about $(1/6) \nu_c^{\rm tri}$.  While it would be desirable to have a
more complete Lindemann analysis in which the collective
modes\cite{shm1add} of these new lattices are quantised, we
believe that the simple analyses presented here capture the essential
physics of the relative stability of the states to quantum
fluctuations.

At filling fractions above $\nu=1/2$ 
quantum fluctuations of the stripes and bubbles
are strongly suppressed. For the cases in
Fig.~\ref{fig:half}(b) and (c), increasing the number of particles to
$N=12$ (so that the respective filling fractions are $\nu= 2/3$ and
$3/4$) leads to much improved groundstate quasi-degeneracies. Quantum
fluctuations are enhanced for $\nu < 1/2$.  
For small non-zero $\alpha$, we find incompressible
liquids at filling fractions, $\nu=p/(3p\pm 1)$, expected for
composite fermions\cite{CooperWadd} formed from bosons bound to {\it
three} vortices.  The state at $\nu=1/4$ is well described by the
$\nu=1/4$ Laughlin state. For $\alpha\agt 1.7$ this state is replaced
by bubble phases, consistent with the expectation from the Lindemann
analysis that for large $q$ these can have critical filling fractions
less than 1/4.

This work was partially supported by the UK EPSRC Grant No. GR/R99027/01 (N.R.C.)
and by the US DOE under contract DE-FG03-02ER-45981 (E.H.R.).

\vskip-0.6cm


\begin{thebibliography}{10}
\bibitem{as} $a_s$  parameterises the short-range
potential and can itself depend on the atomic dipole moment.
\bibitem{symmetry} We cannot rule out lattices of other symmetries,
but believe we have found the important cases for $\alpha \alt 3$.
\bibitem{narrow}
Over the narrow range $\alpha = 0.578-0.600$ the
groundstate is a distorted stripe, of a form intermediate between the
stripe and $q=4$ bubble states.
\end{thebibliography}

\end{document}